\documentclass{article}
\usepackage{epsfig}
\begin{document}

\title{OBSERVATION OF A UNITARY CUSP IN THE THRESHOLD 
${\gamma}p \rightarrow \pi^{0}p$ REACTION}
\author{A. M. Bernstein $^a$, E. Shuster $^a$, R. Beck $^c$, M. Fuchs
$^b$ \\
B. Krusche $^b$, H. Merkel $^c$, H. Str\"{o}her $^b$ \\ 
{\it {\footnotesize $^a$ Physics Department and Laboratory for Nuclear
Science, M.I.T.,}} \\
{\it {\footnotesize Cambridge MA, 02139, USA}} \\
{\it {\footnotesize $^b$ II. Physikalisches Institut, Universit{\"{a}}t
Gie{\ss}en, D-35392 Gie{\ss}en, Germany}} \\
{\it {\footnotesize $^c$ Institut f{\"{u}}r Kernphysik, Johannes-Gutenberg-
Universit{\"{a}}t Mainz}}, \\ 
{\it {\footnotesize D-55099 Mainz, Germany}}}
\maketitle

\begin{abstract}
A rigorous multipole analysis of the recent ${\gamma}p \rightarrow
\pi^{0}p$ cross section measurement is presented. The data were taken
using the photon spectrometer TAPS  at the tagged photon beam of the
Mainz microtron. The s and p wave multipoles were extracted using
minimal model assumptions. The predicted unitary cusp for the s wave
multipole  $E_{0+}$  due to the two step ${\gamma}p \rightarrow
\pi^{+}n\rightarrow \pi^{0}p$ reaction was observed. The results are
consistent with one loop chiral perturbation theory calculations for
which three low energy constants have been determined by a fit to the
data. The uncertainties in the analysis  and the need for polarization
observables are discussed. 
\end {abstract}  

\section{Introduction}

Experiments on photo-pion production from the nucleon are important
because the pion is approximately a Goldstone Boson of QCD
\cite{pi-mass}. The consequences of this are a relatively small mass
(due to the small up and down quark mass) and a weak ${\pi}N$
interaction at low energies \cite{pi-mass}. These characteristics allow
a QCD based approximation scheme known as chiral perturbation theory
(ChPT) \cite{pi-mass,chpt,chpt3}.  Using this technique extensive (1
loop) calculations for threshold photo- and electro-pion production have
been performed \cite{loop}. 

For a long time there has been a predicted unitary 
cusp in the s wave ${\gamma}p \rightarrow \pi^{0}p$ electric dipole 
amplitude, $E_{0+}({\gamma}p \rightarrow \pi^{0}p)$, due to the two 
step ${\gamma}p \rightarrow \pi^{+}n\rightarrow \pi^{0}p$ 
reaction \cite{cusp}. The reason is that the electric dipole
amplitude for  
the  ${\gamma}p \rightarrow \pi^{+}n$ reaction, $E_{0+}( {\gamma}p 
\rightarrow \pi^{+}n)$, is more than an order of magnitude larger 
then $E_{0+}({\gamma}p \rightarrow \pi^{0}p)$ and also because the
threshold energies for the ${\gamma}p \rightarrow \pi^{0}p$ and
$\pi^{+}n$ channels are different (see Table~\ref{tab:thresh}). For this
reason the unitary cusp is isospin violating.

The magnitude of the cusp is related to
\begin{equation}
\label{eq:beta}
\beta=E_{0+}({\gamma}p \rightarrow \pi^{+}n)\cdot a_{cex}(\pi^{+}n
\leftrightarrow \pi^{0}p) 
\end{equation}
where $a_{cex}$ is the s wave charge exchange scattering
length. Recently it was pointed out  
that an accurate measurement of the energy dependence of the 
unitary cusp would allow one to make a measurement of this 
important and previously unmeasured scattering 
length \cite{scat-len}. Furthermore it was also shown that the
two step reaction is expected to exhibit an additional isospin
violation \cite{scat-len} as a consequence of the predicted isospin
violation in $ a_{cex}$ due to the mass difference of the up and down
quarks \cite{iso-quark}. Consequently, experimental tests of the
predictions of ChPT \cite{loop} and the unitary cusp with its light
quark dynamics are of great importance.

The first measurement of the  ${\gamma}p \rightarrow \pi^{0}p$ threshold
cross sections with a 100\% duty cycle electron accelerator was
performed at Mainz \cite{mainz}. The data confirmed a previously
measured total cross section at Saclay \cite{saclay} which was obtained
with a 1\% duty cycle linac and consequently had larger errors. The Mainz
differential cross section qualitatively showed the predicted unitary
cusp for $E_{0+}$ \cite{beck-cusp}. The original interpretation of the
differential cross section data \cite{mainz,saclay} showed a
disagreement with the ``Low Energy Theorems'' (LET)
\cite{let1,let2}. However it was later shown 
\cite{reanalys} that when the total cross section data were included
that the results were consistent with the LET prediction
\cite{let1}. Subsequently it was shown that the 
LET were slowly converging \cite{loop,let2}  and that the prediction for
$E_{0+}$ should be significantly smaller in magnitude. The exact value
is not predicted by ChPT since it depends on low energy constants which
have to be evaluated from experiment \cite{loop}. It was clear that such
an important measurement should be repeated with improved equipment. A
subsequent experiment was performed at Mainz with the TAPS photon
detector \cite{fuchs} where the data from threshold to 152 MeV were
presented after a preliminary analysis.

In this paper a more thorough analysis from threshold to 160 MeV will be
presented. The main purpose of this paper is to obtain the most accurate
values of the multipoles with the minimum number of model dependent
assumptions and to compare these results with the ChPT fit
\cite{loop}. A secondary purpose is to show the model dependence of the
extracted multipoles and the limits due to the fact that the existing
database contains only unpolarized cross sections. The results presented
here are in good agreement, as expected, with our previous publication
\cite{fuchs}. Recently an experiment from Saskatoon has 
been reported \cite{saskatoon} and will also be discussed. 

\section{Formulas and Data Analysis } \label{form-and-anal}

Near threshold one can safely assume that the pions are produced in 
s and p wave states. The differential and total cross sections are:
\begin{equation} \label{eq:sig}
\begin{array}{c}
\sigma(\theta)= q/k [  A + B cos(\theta) + C cos^{2}(\theta)] \\
\sigma_{T}= 4\pi q/k [  A  + C/3]
\end{array}
\end{equation}					
where $q$ and $k$ are the pion and photon center of mass momenta.

It is conventional to compare theory and experiment in terms of
multipole amplitudes. These are: s wave electric dipole $E_{0+}$; p wave
magnetic 
dipole with $j= 1/2$ $M_{1-}$; and  p wave magnetic 
dipole and electric quadrupole amplitudes with $j= 3/2$ $M_{1+}$ and $
E_{1+}$. The A, B,  and C coefficients are quadratic combinations of these
4 amplitudes. Following a previous notation \cite{loop} we
define:
\begin{equation}
\label{eq:P}
\begin{array}{c}
   	P_{1}= 3  E_{1+} + M_{1+} - M_{1-} \\
	P_{2}= 3  E_{1+} - M_{1+} + M_{1-} \\
	P_{3}= 2 M_{1+} + M_{1-} \\
    	{\mid}P_{23}{\mid}^{2}= ({\mid}P_{2}{\mid}^{2} + {\mid} P_{3}{\mid}^{2})/2
\end{array}
\end{equation}

The A, B, and C coefficients are:
\begin{equation}
\label{eq:ABC}
\begin{array}{c}
	A= {\mid}E_{0+}{\mid}^{2} + {\mid}P_{23}{\mid}^{2} \\
	B= 2 Re( E_{0+} P_{1}^{\star}) \\
	C= \mid P_{1}{\mid}^{2} -{\mid}P_{23}{\mid}^{2}
\end{array}
\end{equation}

One can see that  an accurate measurement of $\sigma(\theta)$ for 
unpolarized photons determines 3 linear combinations of the 
multipoles (A, B, and C). On the other hand there are 7 unknown 
parameters, namely the real and imaginary parts of the 4 multipoles 
minus 1 arbitrary overall phase. In the threshold region one can take 
advantage of the fact that the p wave $\pi N$ phase shifts are 
small \cite{hohler} which means that the imaginary parts of the p wave 
multipoles are negligible \cite{loop,scat-len}.

In order to fit the data one must take the predicted unitary cusp in
$E_{0+}$ \cite{loop,cusp,scat-len} into account. This is caused by the
relatively  strong two step ${\gamma}p \rightarrow \pi^{+}n\rightarrow
\pi^{0}p$ reaction channel and a static isospin violating effect
which occurs because of the threshold difference in the  ${\gamma}p
\rightarrow \pi^{0}p$ and  ${\gamma}p \rightarrow \pi^{+}n$ reaction
channels as shown in Table~\ref{tab:thresh}. The first derivations used
a single scattering K matrix approach to calculate the effect of the
final state charge exchange (CEX) \cite{cusp}. The ChPT calculations are
basically isospin conserving but the biggest isospin non-conserving
effect due to the pion mass difference has been included
\cite{loop}. These  approximations can be overcome by using a 3 channel
S matrix  approach in which unitarity and time reversal invariance are satisfied \cite{scat-len}. The resulting equation is:
\begin{equation}
 E_{0+}({\gamma}p \rightarrow \pi^{0}p) = e^{i\delta_{0}  }  [A_{0} + i 
A_{+} a_{cex} q_{+} ]
\end{equation}
where $\delta_{0}$ is the s wave $\pi^{0}p$ phase shift, $A_{0}$ and 
$A_{+}$ are $E_{0+}({\gamma}p \rightarrow \pi^{0}p)$ and $E_{0+}( 
{\gamma}p \rightarrow \pi^{+}n)$  in the absence of the final state 
charge exchange (CEX) reaction, and $q_{+}$ is the $\pi^{+}$ center of 
mass momentum in units of $m_{\pi^{+}}$. For photon energies 
$k_{\gamma} < k_{T}(\pi^{+}n)$, the $\pi^{+}n$ threshold energy, 
one must analytically continue $q_{+}\rightarrow
i{\mid}q_{+}{\mid}$. This switching of the amplitude from real to imaginary
as the secondary threshold opens is the sign of a unitary cusp.
\begin{table} \centering
\begin{tabular}{||l|c||}  \hline\hline
\multicolumn{1}{||c|}{Reaction} & \multicolumn{1}{|c||}{Threshold Energy, MeV} \\
\hline
${\gamma}p \rightarrow \pi^{0}p$ & 144.68 \\ \hline
${\gamma}p \rightarrow \pi^{+}n$ & 151.44 \\ \hline\hline
\end{tabular}
\caption{Threshold energies} \label{tab:thresh}
\end{table}

We can safely neglect $\delta_{0}$  in the threshold region because 
the s wave scattering length $a(\pi^{0}p)$ is expected
\cite{iso-quark,hohler,s-pred} to be very small ($\stackrel{<}{\sim}
0.01/m_{\pi}$). Since the effect of the $\pi^{0}p$ channel is expected
to be small in the $\pi^{+}n$ channel we can take $A_{+}\cong
E_{0+}({\gamma}p \rightarrow \pi^{+}n)$. With these two mild
approximations the 3 channel S matrix formulation reduces to the
previously obtained formulas \cite{loop,cusp}:
\begin{equation}
E_{0+}({\gamma}p \rightarrow \pi^{0}p) =  A_{0}(k_{\gamma}) + i \beta q_{+}
\end{equation}
where the only first principles constraint  that we have for  $A_{0}$ 
is that it is a smooth function of  $k_{\gamma}$. Note that for 
$k_{\gamma} < k_{T}(\pi^{+}n)$, $E_{0+}({\gamma}p \rightarrow 
\pi^{0}p) = A_{0}(k_{\gamma}) - i \beta{\mid}q_{+}\mid$ is purely
real. For $k_{\gamma} > k_{T}(\pi^{+}n), E_{0+}({\gamma}p \rightarrow
\pi^{0}p) $ is complex with $Re E_{0+} = A_{0}(k_{\gamma})$, a smooth
function of $k_{\gamma}$, and $Im E_{0+} = \beta q_{+}$, the cusp
function. The same function $A_{0}(k_{\gamma})$ and parameter $\beta$
occurs both below and above $ k_{T}(\pi^{+}n)$. 

\begin{figure} \centering
\epsfig{file=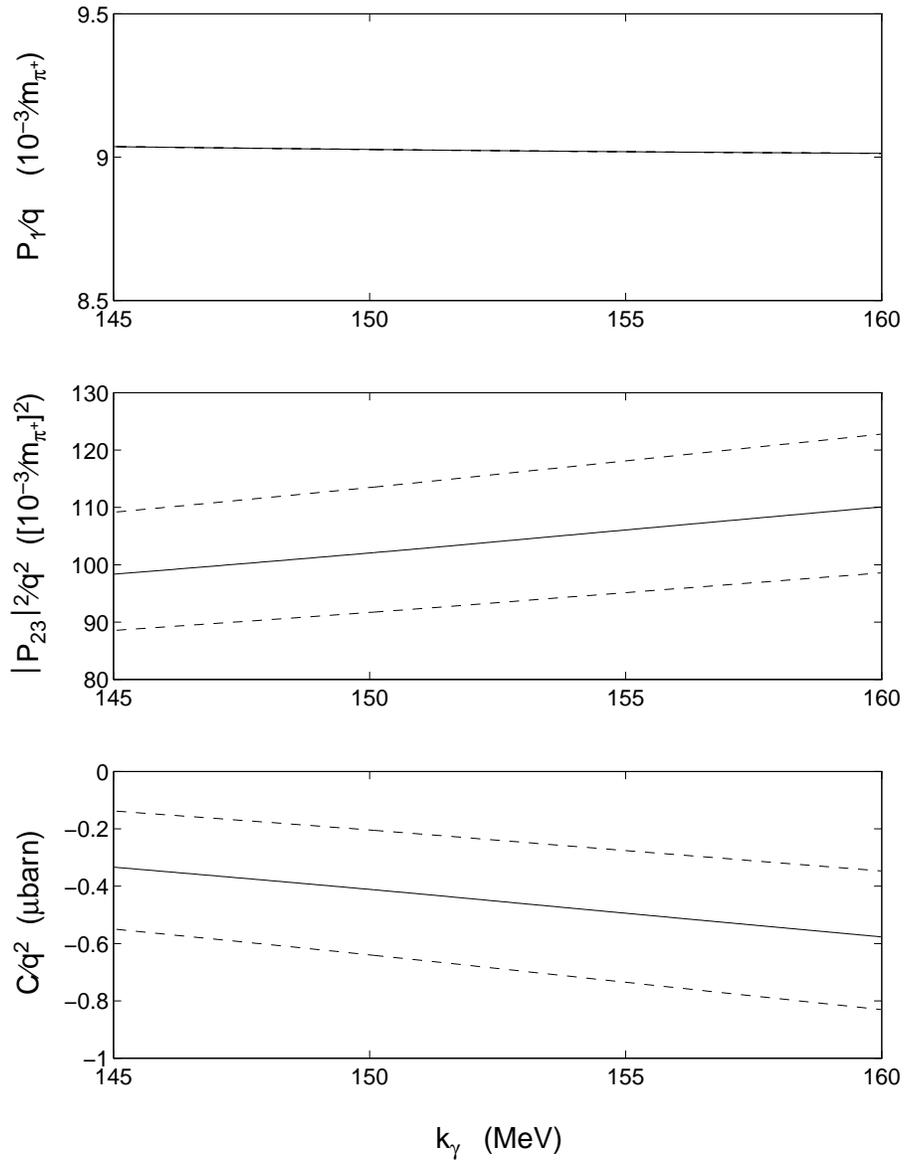,width=340pt}
\caption{ChPT predicted behavior of the p waves. Solid
line represents the predicted values and the dashed lines represent a
$\pm$10\% variation of $b_{p}$.} \label{fig:bp}
\end{figure}

To determine the p wave multipoles we need to consider their energy 
dependence. It was  previously assumed that they go to zero as $qk$ for
$k_{\gamma} \rightarrow k_{T}(\pi^{0}p)$ 
\cite{qkp-wave}; recently it 
has been shown that the factor of $k$ should not be there
\cite{loop}. Numerically the difference is not large but
the proper form will be used here. These threshold arguments alone
cannot determine over what range of $k_{\gamma}$ this simple energy
dependence is expected to be valid. In order to see this we plot in
Fig. \ref{fig:bp} the energy  dependence of the three p wave observables
as predicted by ChPT  \cite{loop} up to $k_\gamma= 160
MeV$. We observe that $P_{1}$ is predicted to be very close to  linear
with q for the entire energy region. 
If all of the p wave multipoles were linear in q then $P_{23}^{2}$ and 
C would be proportional to $q^{2}$. As can be seen in Fig. 1 there is a 
deviation from this quadratic dependence which is approximately linear in 
${\Delta}k_{\gamma}= k_{\gamma}-k_{T}(\pi^{0} p)$. 

The predictions of ChPT for the ${\gamma}N \rightarrow {\pi}N$ reaction
depend on three low energy constants labeled $b_{p}$, $a_{1}$, and
$a_{2}$ \cite{loop}. Of these only $b_{p}$ effects the p wave
multipoles. In Fig. 1 the variation of the three observables are also
shown for a $\pm$10\% variation of $b_{p}$. There is no dependence of
$P_{1}$ on $b_{p}$. Furthermore, the slope of ${\mid}P_{23}\mid$ and C
are approximately independent of $b_{p}$. The approximately linear
deviation from the 
$q^{2}$ dependence of C and ${\mid}P_{23}{\mid}^{2}$, shown in
Fig.~\ref{fig:bp}, will be assumed in the analysis but with empirically
determined constants.

We have performed three fits to the data. One uses  the A, B, and C 
coefficients with the energy dependence of C specified as
\begin{equation} \label{eq:C}
	  C = q^{2} [\overline{C} + \widetilde{C} {\Delta}k_{\gamma}] 
\end{equation}
where ${\Delta}k_{\gamma}$ is taken for convenience to be in units of 
$m_{\pi^{0}}$.

A multipole fit was also performed with the functional form of the
energy dependence of 
the p wave multipoles and $Im E_{0+}$ fixed. From the discussion of 
the expected energy dependence of the s and p wave multipoles the 
following energy dependence was chosen:

\begin{equation} \label{eq:stand}
\begin{array}{c}
\begin{array}{lr}
Im  E_{0+}=  0          &         k<k_{T}(\pi^{+}n) \\
Im  E_{0+}=   \beta q_{+}   &   k>k_{T}(\pi^{+}n)
\end{array}  \\
P_{1} = q \overline{P_{1}} \\
{\mid}P_{23}{\mid}^2 = q^{2} [\overline{P_{23}} +
\widetilde{P_{23}}{\Delta}k_{\gamma} ] 
\end{array}
\end{equation}
where the fit parameters are the values of $Re E_{0+}(k_{i})$ at each 
photon energy, $\beta,\overline{P_{1}}$, $\overline{P_{23}}$,and 
$\widetilde{P_{23}}$. 

We have also performed a unitary fit for which $E_{0+}(k_{\gamma})$ is
parametrized as:

\begin{equation} \label{eq:linear}
\begin{array}{c}
E_{0+}(k_{\gamma})= A_{0}(k_{\gamma}) + i \beta q_{+} \\
A_{0}(k_{\gamma}) = \overline{A}+\widetilde{A}{\Delta}k_{\gamma}
\end{array}
\end{equation}
where the values of $\overline{A}$, $\widetilde{A}$, and $\beta$ were taken
as free parameters.

To calculate the expected value of $\beta$ we use the best experimental
value of $a(\pi^{-}p\rightarrow\pi^{0}n) = -(0.1301\pm 0.0059) /m_{\pi}$
from the observed width in the 1s state of pionic hydrogen atom
\cite{pionH2}. This is in excellent agreement with ChPT predictions of
$-(0.130\pm0.006) /m_{\pi}$ \cite{s-pred}. Assuming isospin is conserved
$a(\pi^{+}n \leftrightarrow \pi^{0}p) = -a(\pi^{-}p \leftrightarrow
\pi^{0}n)$. There are no modern
measurements for  $E_{0+}({\gamma}p \rightarrow \pi^{+}n)$ so we can use
the ChPT prediction of $28.2\pm0.6$ \cite{loop,units,chpt-pred}. From
these one obtains $\beta=3.67\pm0.18$ \cite{units}.

The analysis that will be presented here depends on the range of 
validity of Eq.~(\ref{eq:stand}). In order to insure that this
analysis is accurate and 
as model independent as possible it will be terminated at a photon 
energy of 160 MeV. This is sufficient to show the main features of 
the threshold region.

\section{Results}

In this section the results of the analysis of the data \cite{fuchs}
will be presented and compared with ChPT 
calculations \cite{loop}. The data were taken from threshold to 270 MeV. 
In the first publication only a preliminary analysis of the data to a 
photon  energy of 152 MeV was presented  \cite{fuchs}.  In this 
publication a more thorough  analysis to 160 MeV will be presented. 

\begin{figure} \centering
\epsfig{file=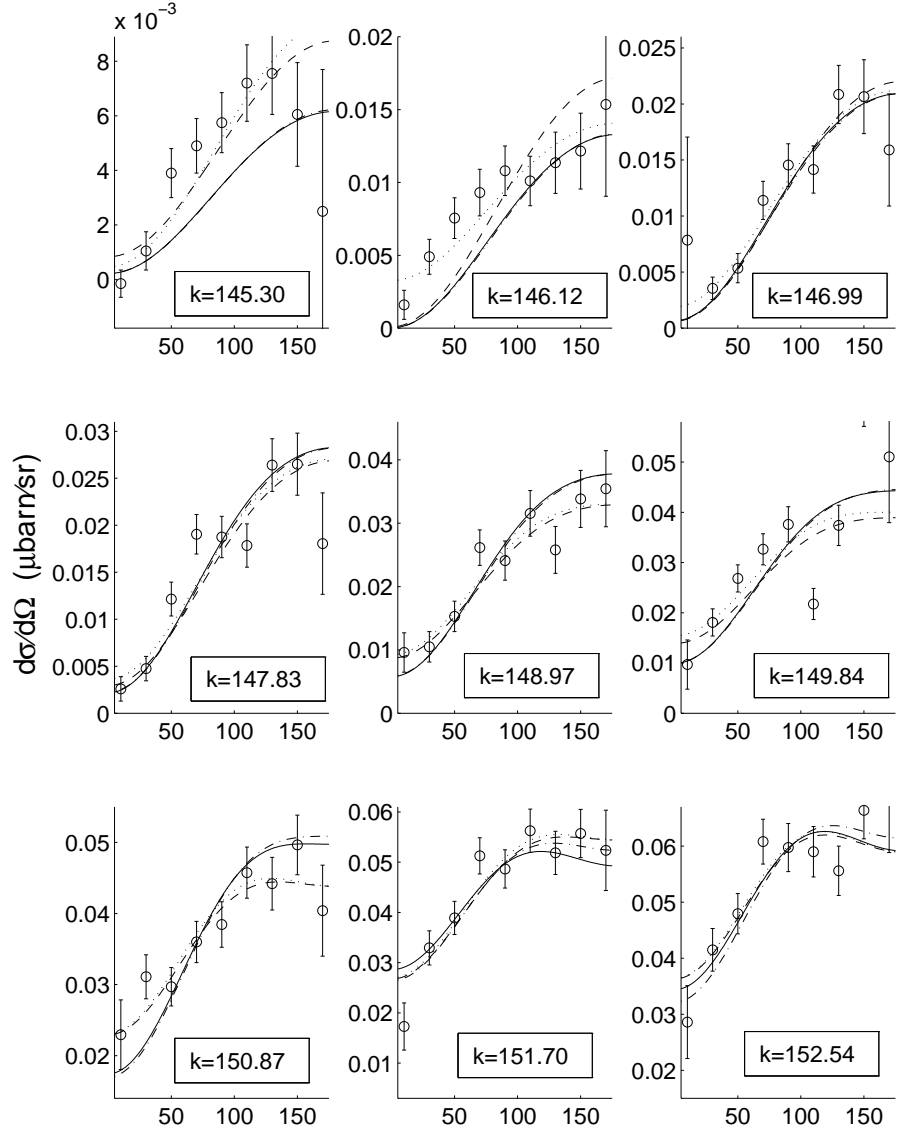,width=340pt}
\caption{Differential cross sections (CM system) with statistical errors
in ${\mu}b/sr$ versus pion angle in degrees. The photon energies in MeV
are given. The solid line represents the unitary fit, dot-dot line the
$A,B,\overline{C},\widetilde{C}$ fit, dash-dash line the multipole fit,
and dash-dot line the ChPT fit.} \label{fig:dsig} 
\end{figure}
\addtocounter{figure}{-1}
\begin{figure} \centering
\epsfig{file=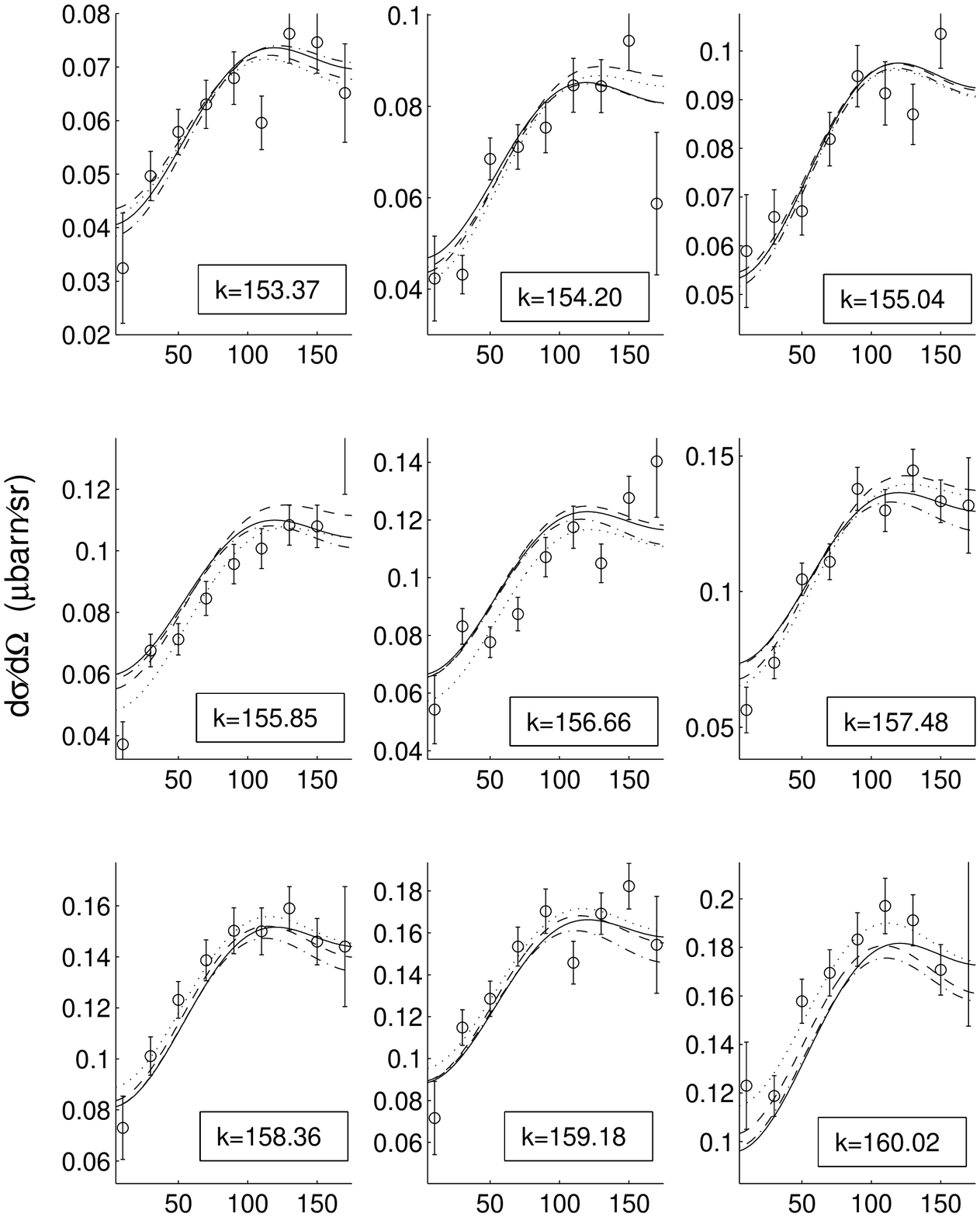,width=345pt}
\caption{Continued.}
\end{figure}


The empirical fits and ChPT \cite{loop} results for $\sigma(\theta)$ are
compared to experiment in Fig.~\ref{fig:dsig}. 
It should be pointed out that the errors shown in Fig.~\ref{fig:dsig}
are statistical only. The systematic errors are estimated to be
approximately 5\% \cite{fuchs}. For the A,B,C fit (Eq.~(\ref{eq:C})) the
best fit values are $\overline{C}=-0.338 \pm 0.025 {\mu}b$,
$\widetilde{C}=-1.83 \pm 0.30 {\mu}b$, and $\chi^{2}/DOF=1.65$. For ChPT
\cite{loop} there are three low energy constants which were adjusted to
obtain a best fit to the data. Qualitatively all of the fits and the
ChPT calculation are very close with $\chi^{2}/DOF= 1.96$ for the
multipole fit and 2.21 for ChPT.

The fit parameters and $\chi^{2}$ results are shown in
Table~\ref{tab:pars} and compared to ChPT. Only the fitting errors are
presented. The values presented for ChPT \cite{loop} were obtained by
fitting the numerical calculations with Eqs.~(\ref{eq:stand}) and
(\ref{eq:linear}) and finding the best fit parameters. The agreement
between the fitted form and numerical results is excellent. 

For the p wave multipoles one can see from Table~\ref{tab:pars} that the
extracted values of $\overline{P_{1}}$ and $\overline{P_{23}}$ are in
good agreement with ChPT \cite{loop}. In this part of the calculation
there is only one low energy constant ($b_{p}$) which was fit to the
data. On the other hand it can be seen that there is a large systematic
error for $\widetilde{P_{23}}$. This can  be seen by comparing the much
different values obtained from the unitary and multipole fits. These
results straddle the ChPT value. The fact that the next to leading order
slope of the p wave multipoles is not strongly constrained by the data
will have consequences in the determination of $Im E_{0+}$ (see
Sec.~\ref{mod-dep}).

The parameters for $Re E_{0+}$ ($\beta$, $\overline{A}$, and
$\widetilde{A}$) are significantly different between the unitary fit and
the ChPT calculation. As will be shown below (see Fig.~\ref{fig:ReE0}
and discussion) the two resulting curves both are in reasonable
agreement with the extracted values of $Re E_{0+}$ from the multipole fit. 
We should point out that  for this multipole there are two low 
energy parameters ($a_{1}$ and $a_{2}$) that have been fit to the data. 

\begin{table} \centering
\begin{tabular}{||l|c|c|c||}  \hline\hline
\multicolumn{1}{||c|}{Parameter} & Unitary Fit & Multipole Fit & ChPT \\ \hline
\rule{0in}{.4cm} $\chi^{2}/DOF$ & 2.13 & 1.96 & 2.21 \\ \hline
\rule{0in}{.4cm} $\beta$ & $3.76 \pm 0.11$ & $2.82 \pm 0.32$ & $2.78$ \\ \hline
\rule{0in}{.4cm} $\overline{P_{1}}$ & $9.006 \pm 0.079$ & $9.151 \pm 0.071$ & $8.998$ \\ \hline 
\rule{0in}{.4cm} $\overline{P_{23}}$ & $101.9 \pm 1.1$ & $95.40 \pm 0.62$ & $97.71$ \\ \hline 
\rule{0in}{.4cm} $\widetilde{P_{23}}$ & $41.8 \pm 13.0$ & $159.1 \pm 7.3$ & $108.5$ \\ \hline
\rule{0in}{.4cm} $\overline{A}$ & $-0.12 \pm 0.020$ & & -0.41 \\ \hline 
\rule{0in}{.4cm} $\widetilde{A}$ & $-4.27 \pm 0.28$ & & -0.76 \\ \hline\hline
\end{tabular}
\caption{Parameters in Eq.~(\protect\ref{eq:stand}) and
(\protect\ref{eq:linear}) for the Unitary and multipole fits and ChPT
\protect\cite{loop}. Except for $\chi^{2}$ all of the quoted parameters
are given in units of $10^{-3}/m_{\pi^{+}}$. In Eqs.~(\protect\ref{eq:stand})
and (\protect\ref{eq:linear}) $q$ and ${\Delta}k_{\gamma}$ are in units of
$m_{\pi^{0}}$ and $q_{+}$ in units of $m_{\pi^{+}}$} \label{tab:pars}
\end{table}

\begin{figure} \centering
\epsfig{file=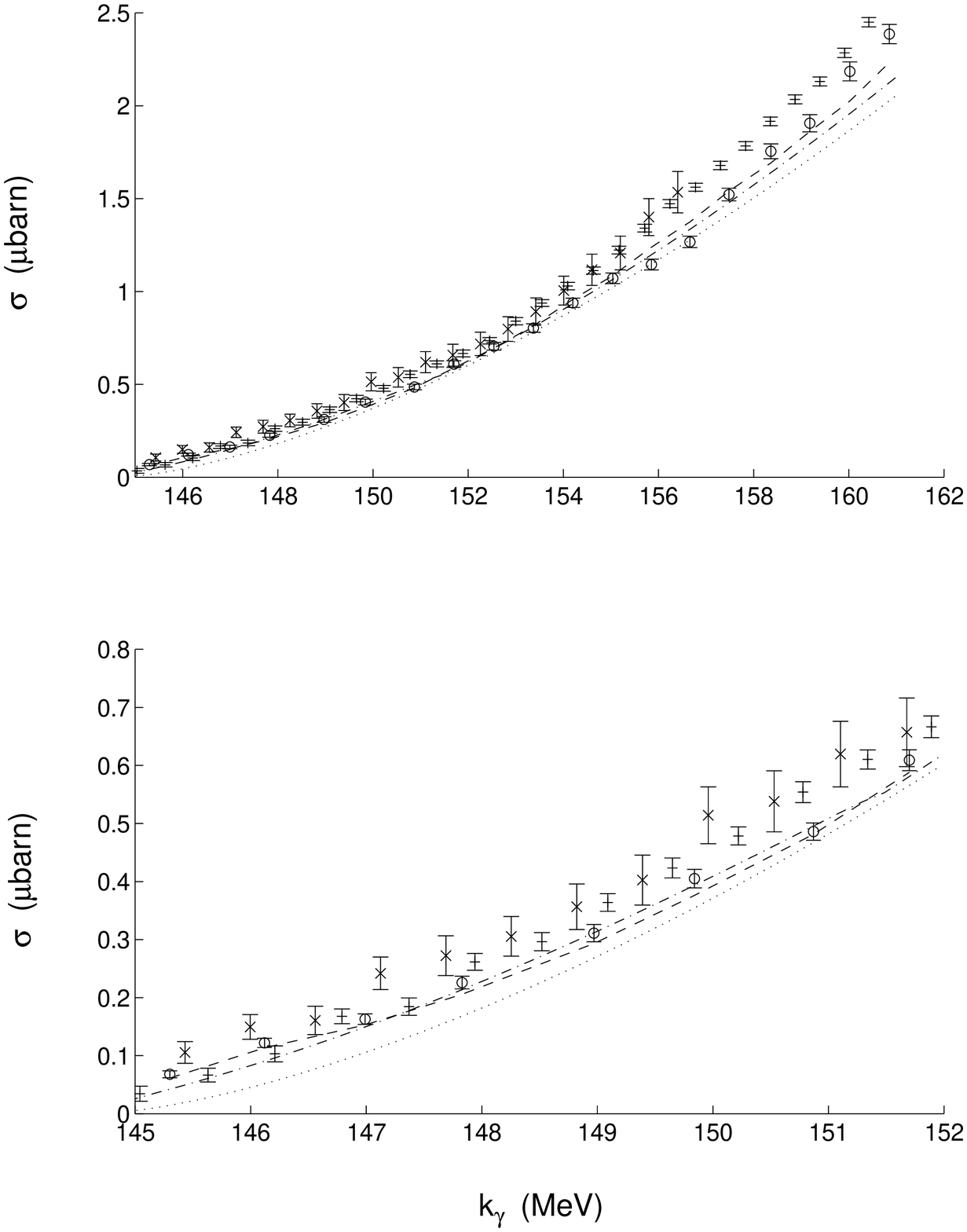,width=345pt}
\caption{Measured total cross section with statistical errors in
${\mu}b$ versus photon energy 
with the ChPT calculations \protect\cite{loop} and the empirical
fits. The new Mainz (TAPS) data (circle) \protect\cite{fuchs}, Saskatoon
(plus) \protect\cite{saskatoon}, and older Mainz points (cross)
\protect\cite{mainz} are shown. The bottom figure shows the energy
region through the $\pi^{+}n$ threshold in more detail. In both figures
dash-dash line represents the multipole fit, dash-dot line the ChPT
result, and dot-dot line only the p wave contribution to the ChPT.}
\label{fig:sigt}
\end{figure}

The results for the total cross section are shown in
Fig.~\ref{fig:sigt}. It can be 
seen that there is a  discrepancy between the Saskatoon \cite{saskatoon} and 
TAPS \cite{fuchs} results particularly for $k_{\gamma} > 152$ MeV. The 
older Mainz results \cite{mainz} tend to be in better agreement with the 
Saskatoon data. Unfortunately at this time the cause of this 
disagreement is not known.

The Saskatoon group did not publish any differential cross section 
data. The only information that is presented is the ``belt pattern'' for 
the angular distributions. In order to use this information one has to 
perform a Monte Carlo calculation using predicted differential cross 
sections which are then compared with the ``belt patterns''. We 
therefore cannot compare the results presented here for the 
differential cross sections with the Saskatoon data.

From Fig.~\ref{fig:sigt} one can see that the ChPT fit
is in good agreement with the TAPS data and not in agreement with the
Saskatoon data. This is not surprising since the three free
parameters of ChPT were fit to the TAPS data. One also notes that the p
wave contribution to the total cross section is dominant except for the
first few points above threshold. This makes it difficult to see the
effect of the (s wave) unitary cusp in the cross section.

\begin{figure} \centering
\epsfig{file=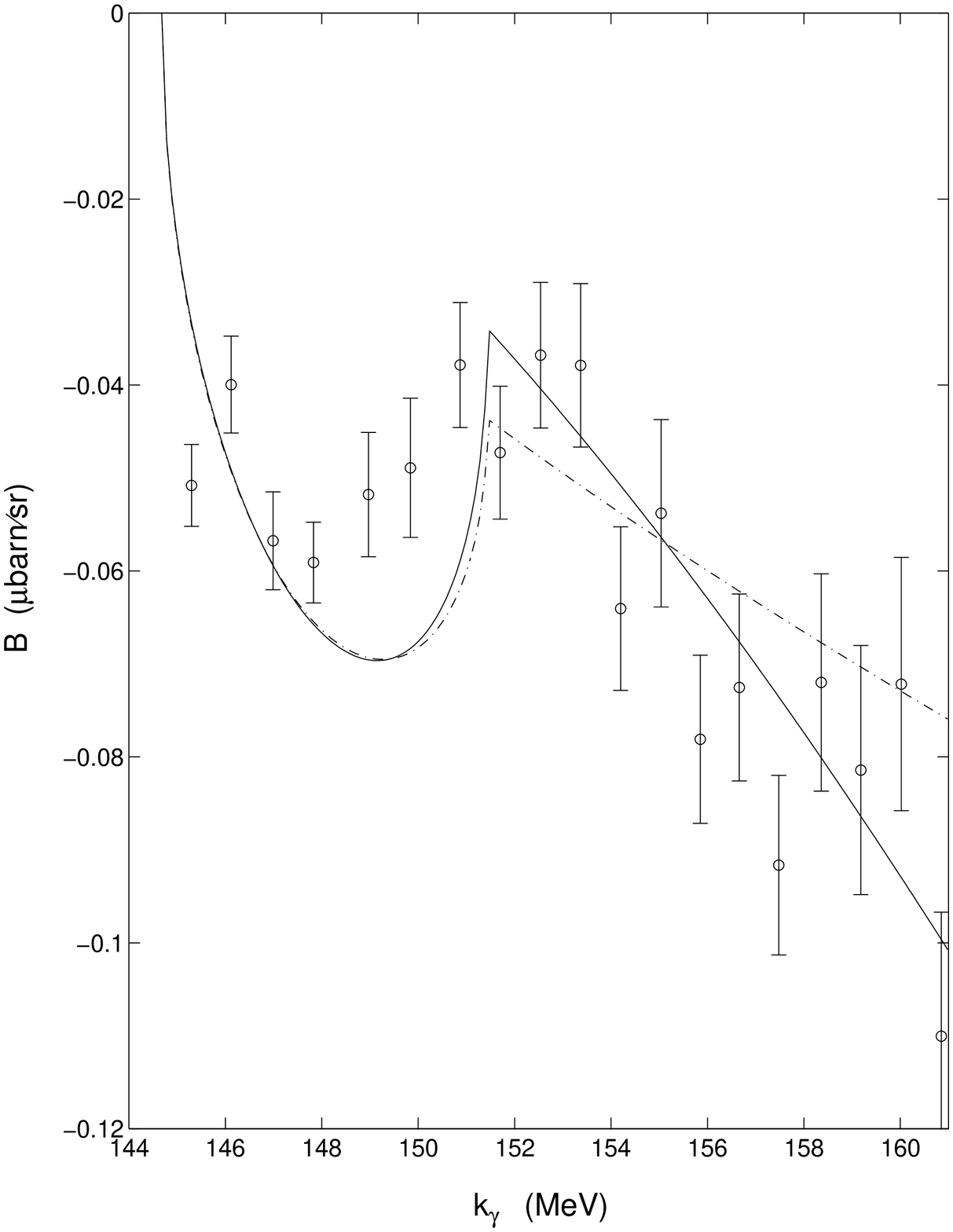,width=290pt}
\caption{B coefficient (with statistical errors) vs. photon energy from
fits and ChPT. Circles represent the $A,B,\overline{C}$ fit, solid line
represents the unitary fit, and the dash-dot line the ChPT.} \label{fig:B}
\end{figure}

The results for the B coefficient are shown in Fig.~\ref{fig:B}. This
shows the effect of the predicted unitary cusp because it is an sp
interference amplitude. The extracted  values of the
B coefficient are in reasonable agreement with the unitary fit and with
ChPT \cite{loop} because only the statistical errors are shown. An
estimate of the systematic errors can be inferred from the scatter of
the points.

\begin{figure} \centering
\epsfig{file=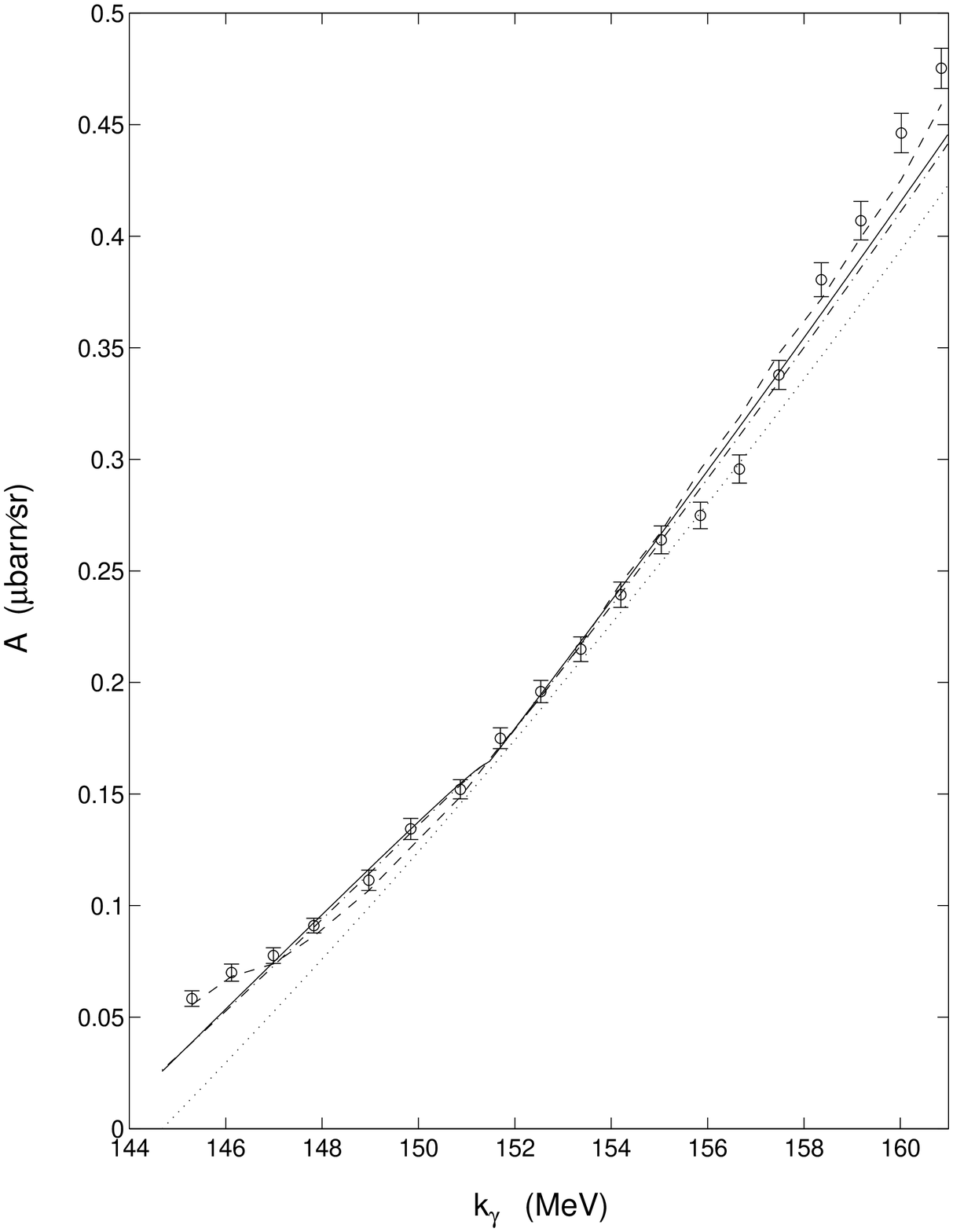,width=290pt}
\caption{A coefficient (with statistical errors) vs. photon energy from
fits and ChPT. Circles represent the $A,B,\overline{C}$ fit, solid line
represents the unitary fit, dash-dash line the multipole fit, dash-dot
line the ChPT, and dot-dot line the p wave contribution to ChPT.} \label{fig:A}
\end{figure}

The results for the A coefficient are shown in Fig.~\ref{fig:A}. This is
the most accurately measured coefficient since $\sigma(\theta=90^{\circ})
=(q/k) A$ and also the total cross section is proportional to $A+C/3$ and
${\mid}C{\mid}\ll A$. As a consequence the errors for A are small. It can be seen, in
contrast to the B coefficient, that the unitary cusp is hardly
visible. This is due to the dominance of p waves as shown in
Fig.~\ref{fig:A}. It can also be seen that  ChPT calculation is in
reasonable agreement with A.
\begin{figure} \centering
\epsfig{file=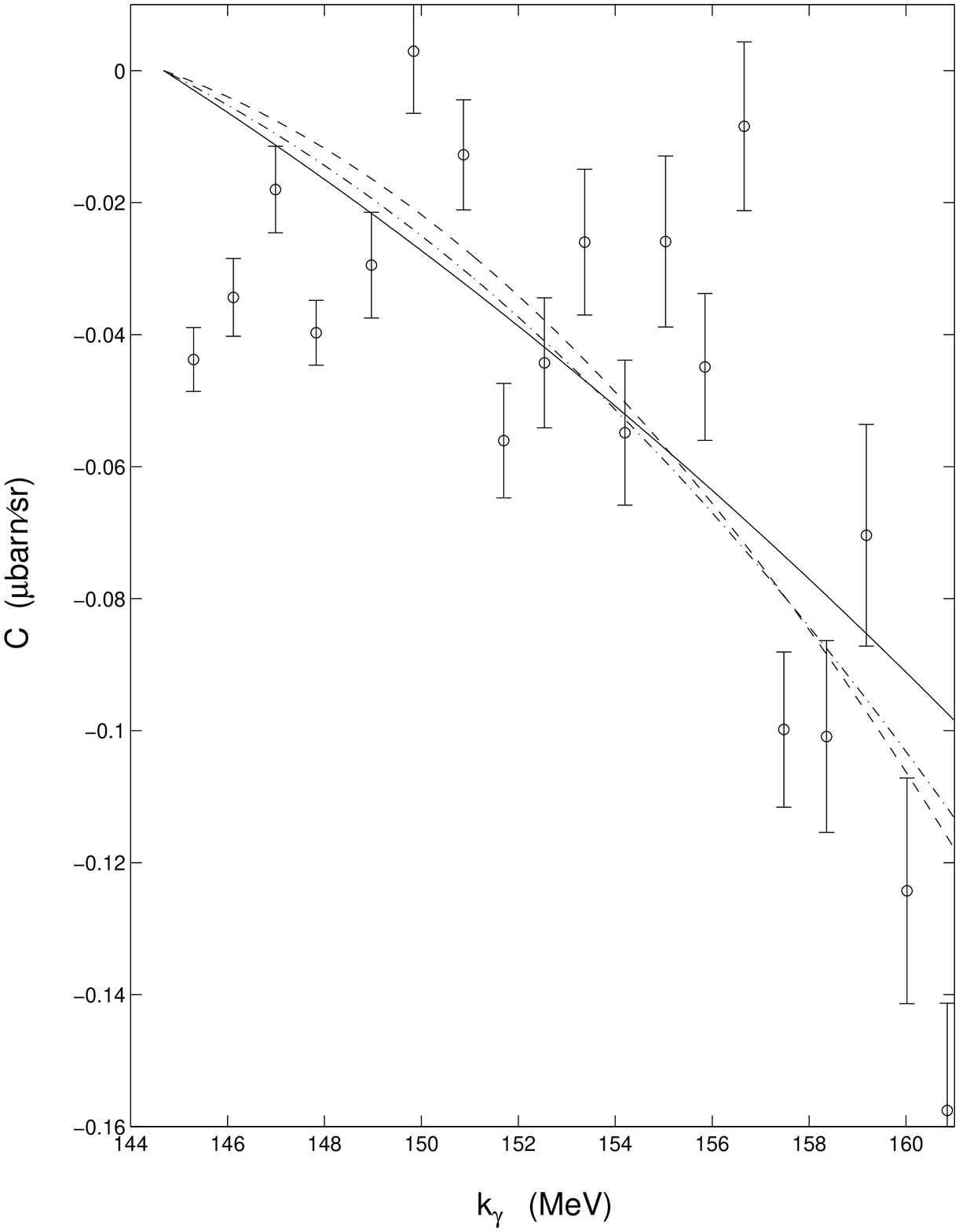,width=290pt}
\caption{C coefficient (with statistical errors) vs. photon energy from
fits and ChPT. Circles represent the $A,B,C$ fit, solid line represents
the unitary fit, dash-dash line the multipole fit, and dash-dot line the ChPT.}
\label{fig:C}
\end{figure}

The results for the C coefficient are shown in Fig.~\ref{fig:C}. It can
be seen that all of the fits and the results of ChPT are in good
agreement. In addition a fit in which the A, B, and C coefficients at
each energy are found from a least square fit to the data are also
presented. In this case the energy dependence of the C coefficient is not
constrained by theory. The scatter of these C coefficients indicate that
the data do not strongly constrain the p wave multipoles despite the
fact that this is the dominant multipole in the unpolarized cross
section. We conclude that although the present results are consistent
with the ChPT theory calculations that the experiment with linearly
polarized photons recently completed at Mainz \cite{beck} is  needed for
a more precise measurement of the p wave multipoles. The results from
the analysis of this experiment will  be available in the next year or
two.

\begin{figure} \centering
\epsfig{file=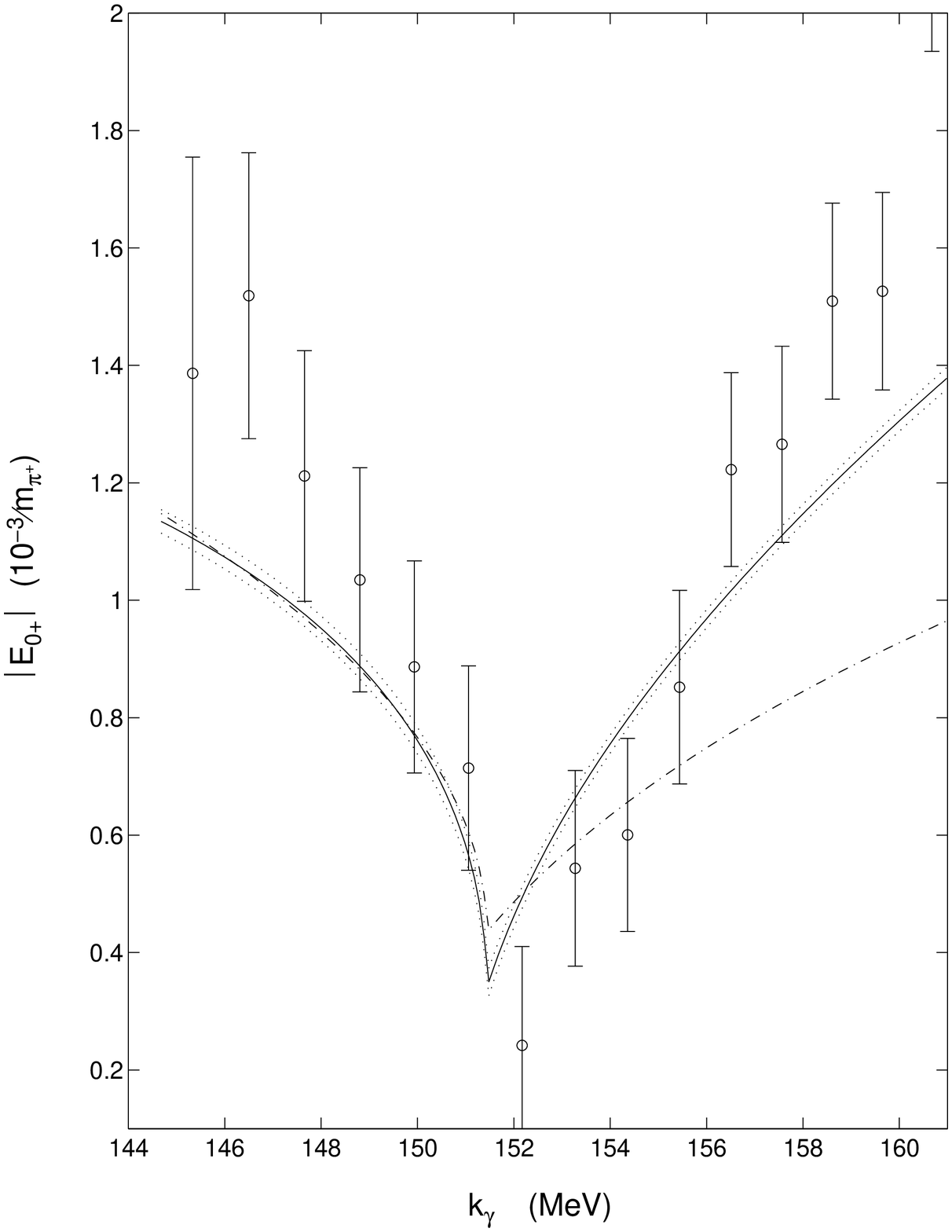,width=290pt}
\caption{${\mid}E_{0+}{\mid}$ vs. photon energy. Circles represent the
Saskatoon data, solid line represents the unitary fit, dash-dot line the
ChPT, and dot-dot lines represent the fitting errors of the unitary fit.}
\label{fig:absE0}
\end{figure}

\begin{figure} \centering
\epsfig{file=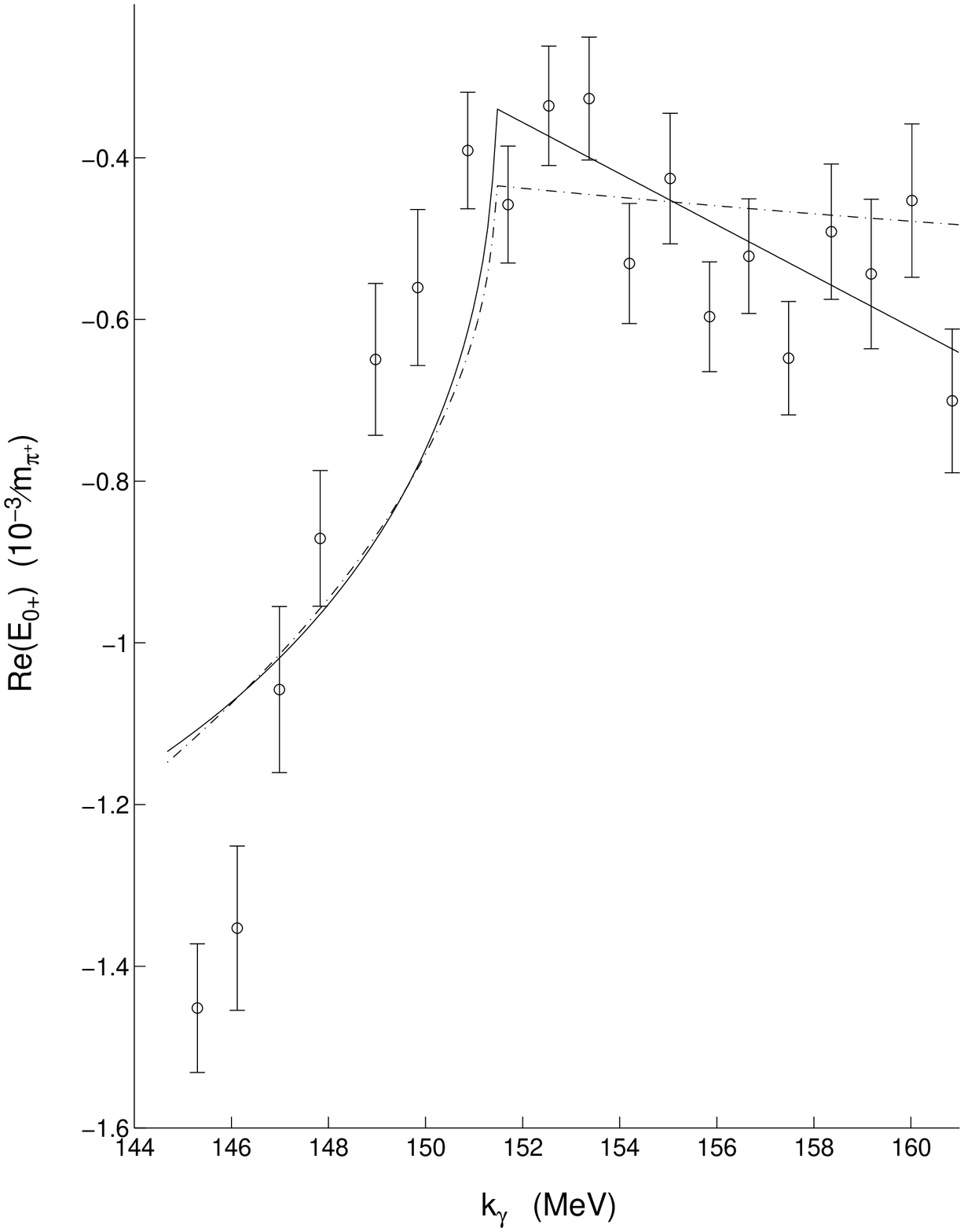,width=290pt}
\caption{$Re E_{0+}$ vs. photon energy. The circles represent the
multipole fit, the solid line represents the unitary fit, and the
dash-dot line the ChPT fit.} \label{fig:ReE0}
\end{figure}

The extracted values of the magnitude of 
\[ {\mid}E_{0+}{\mid}= \sqrt{(ReE_{0+})^{2}+ (Im E_{0+})^{2}} \]
are shown in Fig.~\ref{fig:absE0}. These are obtained from the unitary
fit to the Mainz/TAPS data \cite{fuchs} and from the published Saskatoon
results \cite{saskatoon}. Despite the discrepancy in the measured total
cross sections, the two data sets result in values of ${\mid}E_{0+}{\mid}$
which are in reasonable agreement. The fitting errors for the unitary
fit are significantly smaller than the errors shown for the individual
points of the Saskatoon data since they represent an overall fit to the
data.

The extrapolated threshold values for $E_{0+}$ are $-1.13 \pm 0.04$ for
the unitary fit and $-1.5 \pm 0.1$ for the multipole fit in
the usual units \cite{units}. These two values do not agree because the
two analyses give a somewhat different energy dependence for
$E_{0+}$. If we take the average and adjust the error to reflect this
disagreement the threshold value of $E_{0+} = -1.3 \pm 0.2$
\cite{units}. This is good agreement with the Saskatoon result of $-1.32
\pm 0.1$ \cite{units}. This disagrees with the older prediction of the
low energy theorems \cite{let1,let2} of -2.28 \cite{units} which have
been shown to be incomplete \cite{loop,let2}.

The $Re E_{0+}$ as extracted from the multipole fit is presented in
Fig.~\ref{fig:ReE0}. This agrees with the Saskatoon results
\cite{saskatoon} and the older Mainz results \cite{mainz} within the
experimental errors. The effect of the rapid energy variation of $Re
E_{0+}$ below the $\pi^{+}n$ threshold is again visible in qualitative
agreement with the ChPT calculation \cite{loop} and the unitary
fit. Note that the errors shown in Fig.~\ref{fig:ReE0} are statistical
only. The magnitude of the systematic errors can be inferred from the
scatter of the points.

\section{Model Dependence} \label{mod-dep}

The differential cross section for unpolarized photons in
Eq.~(\ref{eq:sig}) shows that three independent combinations of the
multipoles (A, B, and C) are measured while there are seven independent
multipole parameters for the emission of s and p wave pions. In order to
extract useful information about the multipoles from the data
assumptions were made which follow from first principles which should
cause relatively small analysis errors.  In this section the sensitivity
to these assumptions will be explored to obtain a measure of the
uncertainties, and also to explore how future data, particularly with
polarization degrees of freedom, can improve the situation.

\begin{figure} \centering
\epsfig{file=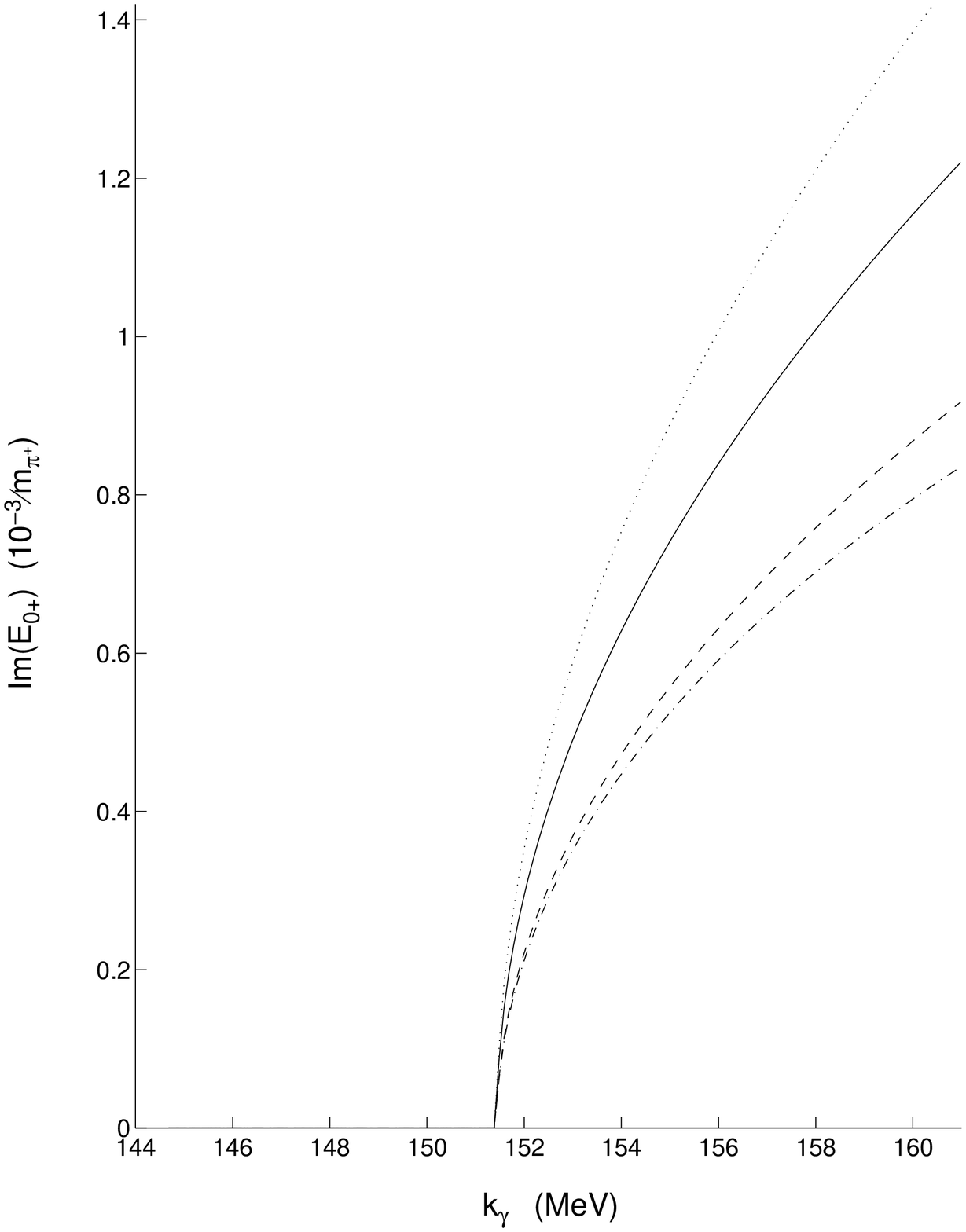,width=290pt}
\caption{$Im E_{0+}$ vs. photon energy. Solid line represents the
unitary fit, dash-dash line the multipole fit, dash-dot line the ChPT,
and dash-dash line the fit which assumes linear dependence of p waves of
$q$.} \label{fig:ImE0}
\end{figure}

The major model assumption that was made in this analysis is the
assumption that the p wave multipoles have the same analytic energy
dependence as predicted by ChPT \cite{loop}. In part this assumption has
been checked by the fact that the least square parameters are close to
those of ChPT (see Table~\ref{tab:pars}). In order to further check this
assumption an additional fit was made which was very similar to the
multipole fit (Eq.~\ref{eq:stand}) except that  the p wave multipoles
were assumed to vary linearly with $q$ (i.e. $\widetilde{P_{23}} = 0$ in
Eq.~\ref{eq:stand}). This fit is quantitatively similar to those in
which the p wave multipoles were assumed to vary as $qk$
\cite{reanalys}. The results of doing this can be surmised by the
observation that A is the best measured of the three coefficients in
Eq.~(\ref{eq:sig}) and by noting that  this determines the absolute value
of $E_{0+}$ in addition to the dominant p wave contribution. Since $Re
E_{0+}$ is determined from the B coefficient this determines $Im E_{0+}$
after a suitable subtraction of the p wave contribution. If one assumes
a smaller energy dependence in the p wave multipole then a stronger
energy dependence will emerge for $Im E_{0+}$. The results for the
determination of $Im E_{0+}$ are presented in Fig.~\ref{fig:ImE0} and it
can be seen that this is precisely what has happened. For the fit in
which the p waves are assumed to be proportional to $q$ the extracted
value of $\beta = 4.51 \pm 0.20$, which is far larger then the value of
$\beta = 2.82 \pm 0.32$ obtained with the multipole fit or
$\beta = 3.76 \pm 0.11$ for the unitary fit. The two latter fits use the
p wave energy dependence of Eq.~(\ref{eq:stand}). These results for
$\beta$ indicate a strong correlation between $\beta$ and
$\widetilde{P_{23}}$. This is to be expected since the only information
about these parameters is obtained from A (see Eqs.~\ref{eq:ABC} and
\ref{eq:stand}).

From the spread in the values of $\beta$ shown in Fig.~\ref{fig:ImE0}
and Table~\ref{tab:pars} one concludes that the systematic error 
is significantly larger then the fit error. This is due to the fact that
there is not 
sufficient information in the unpolarized cross section to determine
this quantity. Experiments with polarized targets are required to
precisely determine $Im E_{0+}$ \cite{scat-len}.

The cusp effect is isospin breaking
due in part to the threshold difference between the $\pi^{0}p$
and $\pi^{+}n$ channels. In the ChPT calculation \cite{loop}
isospin is 
broken by inserting the mass difference between the charged and neutron
pions by hand. This leads to a value of $\beta= 2.78$ which is
significantly below the predicted value of $3.67\pm0.19$ expected from
the predicted values of $E_{0+}({\gamma}p \rightarrow \pi^{+}n)$ and
$a(\pi^{+}n\leftrightarrow\pi^{0}p)$ quoted in
Sec.~\ref{form-and-anal}. An improved ChPT calculation which takes
isospin breaking into account in a more dynamic way seems to be
required to obtain detailed agreement with experiment.

\begin{figure} \centering
\epsfig{file=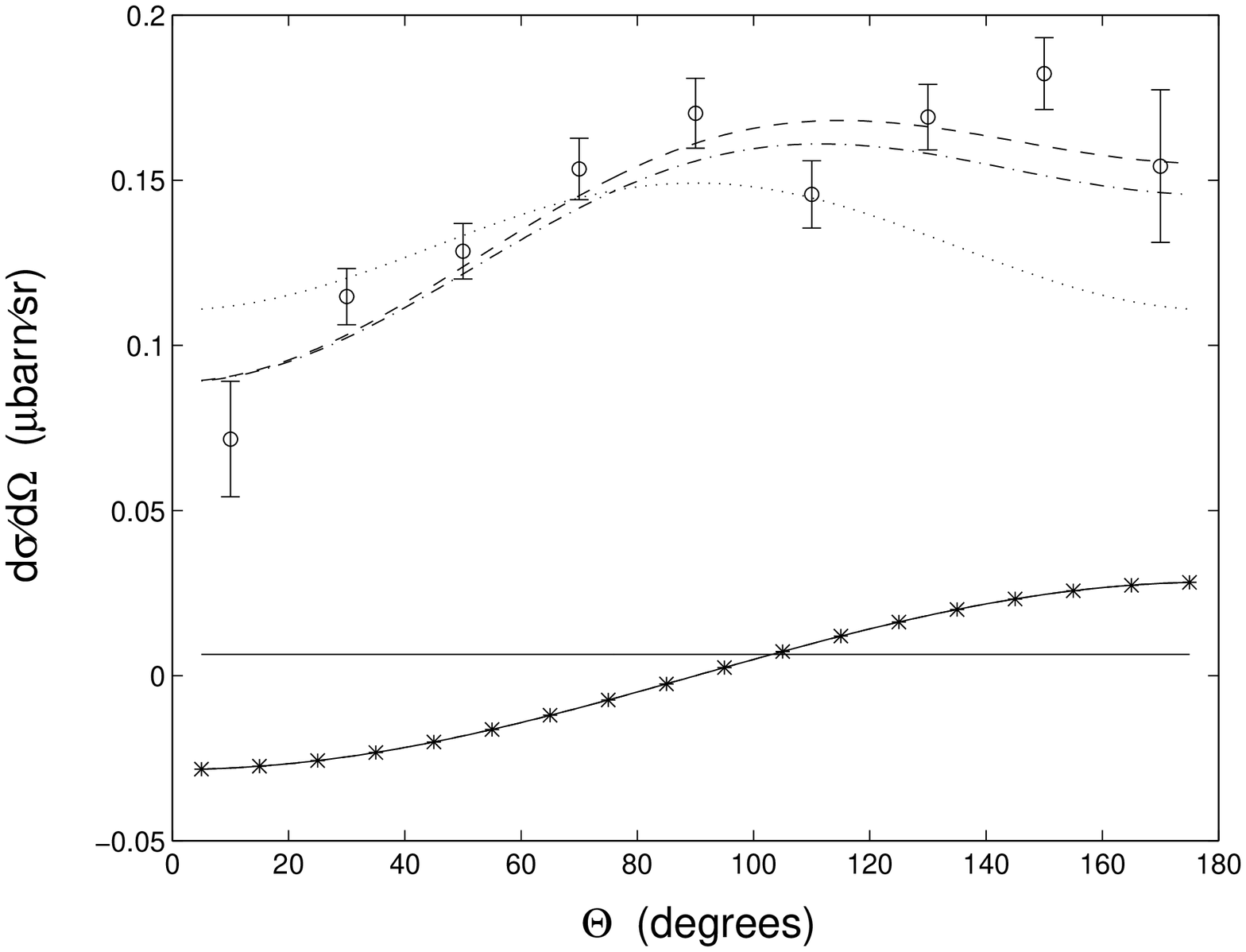,width=280pt}
\epsfig{file=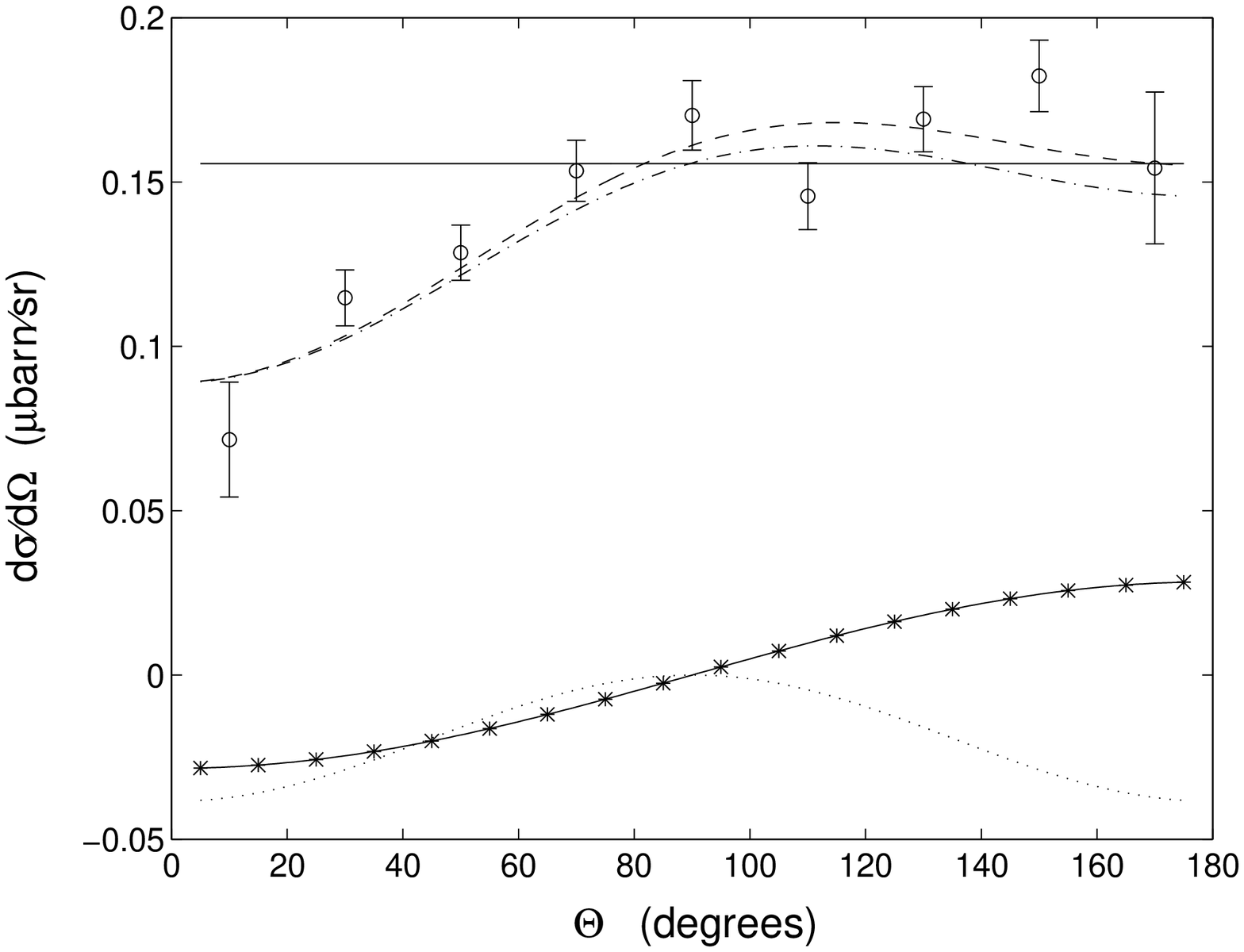,width=280pt}
\caption{$\sigma(\theta)$ at 159.1 MeV showing the contributions of the
A, B, and C terms (bottom) and s and p terms (top). In both graphs the
circles represent Mainz/TAPS data, the dash-dash line represents the
multipole fit, and the dash-dot line the ChPT fit. Solid line represents
the s wave (top) and A term (bottom) contributions, the dot-dot line p
wave (top) and C term (bottom) contributions, and stars represent sp
(top) and B term (bottom) contributions to $\sigma(\theta)$.}
\label{fig:sig159}
\end{figure}

It is also of interest to discuss how well the p wave multipoles are
measured. By comparing the values obtained from the unitary and multipole
fits (Table~\ref{tab:pars}) it can be shown that the leading terms in
the energy dependence of the p wave multipoles ($\overline{P_{1}}$ and
$\overline{P_{23}}$ in Eq.~(\ref{eq:stand})) are well determined, but
not the next order term ($\widetilde{P_{23}}$ in
Eq.~(\ref{eq:stand})). This is a relatively small effect in the
magnitude of the p waves as can be seen from the three curves for C in 
Fig.~\ref{fig:C}. The question of how accurately the energy 
dependence of the p wave multipoles can be obtained from the data can be
seen by examining the relative contributions of the different
contributions (Eqs.~\ref{eq:sig}-\ref{eq:ABC}) to the differential cross
section. As shown in Fig.~\ref{fig:sig159} the p wave multipoles are
dominant. However to extract their precise magnitude from the data is
not trivial. First, as was discussed above, the best measured A
coefficient has a tradeoff between $Im E_{0+}$ and the energy dependence
of the p wave multipoles. The B coefficient is an interference term
between $P_{1}$ and $Re E_{0+}$ so that only the product is
determined. Only the relatively small C coefficient has  purely p wave
contributions. As was shown in the discussion of Fig.~\ref{fig:C} this
coefficient is poorly determined. The reason for this is illustrated in
Fig.~\ref{fig:sig159} which shows the contributions of the A, B, and C
terms to $\sigma(\theta)$ at 159.1 MeV. This is a relatively high energy
for this analysis so that the contribution of the C term is as large as
possible. It can be seen that the C term  contribution is relatively
hard to determine with precision. It will be much better determined by
the $\vec{\gamma} + p \rightarrow \pi^{0} + p$ reaction with linearly
polarized photons \cite{beck}. In this case if one assumes that the
$E_{1+}$ multipole is negligible the polarized photon asymmetry  at
$90^{\circ}$ is $-C/A$. This will be the most precise measurement of the
energy dependence of the p wave multipoles.

\section{Conclusions}

We have presented a rigorous analysis of the recent TAPS/Mainz
data \cite{fuchs}. Since there are three independent observables for the
unpolarized cross section, while for  s and p wave photo-pion emission
there are seven independent amplitudes, some simplifying assumptions
must be made. In this work these assumptions follow closely from first
principles. The systematic errors of the analysis were assessed by using
several assumptions; the main one is that the energy dependence of the p
wave multipoles follow approximately the same analytic form predicted by
ChPT \cite{loop}.

The main conclusion is that the calculation of ChPT \cite{loop} are in
reasonable agreement with the data. This one loop calculation has three
low energy constants which are fitted to the data. The threshold value
of $E_{0+}$ was determined to be $-1.3 \pm 0.2$ in the usual units
\cite{units}. This is in agreement with the Saskatoon result of $-1.32
\pm 0.1$ \cite{saskatoon}. Note that this disagrees with the predictions
of the older "low energy theorems" of $-2.28$ \cite{let1,let2} which have been
theoretically shown to be incomplete \cite{loop}.

The predicted unitary cusp in the s wave electric dipole amplitude
$E_{0+}$ \cite{loop,cusp,scat-len}, due to the two step ${\gamma}p
\rightarrow \pi^{+}n \rightarrow \pi^{0}p$, has been observed in
experiments at Mainz \cite{mainz,beck-cusp,fuchs} and Saskatoon
\cite{saskatoon,saskatoon-new}. Only a range of $\beta$ values
(Eq.~(\ref{eq:beta})) between approximately 2.8 and 4.5 \cite{loop} an
be obtained from the unpolarized cross section data (Sec.~\ref{mod-dep}).

The new experiments on the threshold ${\gamma}p \rightarrow \pi^{0}p$
reaction mark an important advance in our understanding of this
important reaction. An understanding of the small discrepancy between
the TAPS \cite{fuchs} and Saskatoon \cite{saskatoon} results is
important. A new experiment with linearly polarized photons has been
performed \cite{beck} and we look forward to the completion of the data
analysis to more completely test the ChPT predictions for the p wave
multipoles \cite{loop}. Further work with polarized targets is required
to precisely measure $\beta$, which should provide a measure of the s
wave charge exchange scattering length $a_{cex}(\pi^{+}n \leftrightarrow
\pi^{0}p)$ \cite{scat-len}. On the theoretical side we have seen tremendous
progress in our understanding of threshold pion photo and electro
production. Further advances are needed to treat isospin breaking in a
better way and to test the convergence of $E_{0+}$.


\begin{thebibliography}{99}
\bibitem{units} The values  of the multipoles are quoted in the 
traditional units of $10^{-3}/m_{\pi^{+}}$.
\end{thebibliography}

\end{document}